\documentstyle[12pt]{article}
\begin{document}
\title{The Absolute Upper Bound on the 1-loop Corrected mass
of $S_1$ in the NMSSM}
\author{Seung Woo Ham and Sun Kun Oh
        \\{\it Institute for Advanced Physics, Department of Physics,} \\
{\it Kon-Kuk University,} \\
          {\it Seoul 143-701, Korea}
\\
\\
        Bjong Ro Kim
\\
{\it III. Phys. Inst. A, RWTH Aachen,} \\ {\it 52056 Aachen, Germany}
\\
\\
}
\date{}
\maketitle
\begin{abstract}
We examine in detail radiative corrections to the lightest scalar Higgs
boson mass due to the top quark and scalar quark loops in the
next-to-minimal supersymmetric standard model (NMSSM).
We take into account the nondegenerate state for the top scalar quark
masses.
In our analysis, the mass matrix of the top scalar quark contains
the gauge terms.
Therefore our formula for the scalar Higgs boson mass matrix at the 1-loop
level includes the contribution of the gauge sector as well as the effect
of the top scalar quark mass splitting.
Thus we calculate the upper bound on the lightest scalar Higgs boson mass
using our formula.
We find that the absolute upper bound on the 1-loop corrected mass of
the lightest scalar Higgs boson is about 156 GeV.
\end{abstract}
\vfil

\section{INTRODUCTION}

\hspace*{6.mm}
One of the main motivations for a supersymmetric extension of the
standard model (SM) is the fact that a calculation of the 1-loop
corrections to the Higgs boson mass yields a quadratic divergence
arising from the SM particle loops.
In a supersymmetric theory, all particles in the SM are accompanied
by their superpartners.
Therefore, all quadratic divergences are eliminated by the cancellation
between ordinary particle and its superpartner loops.

As a matter of fact, supersymmetry (SUSY) requires the existance of
supermultiplets made up of fermions and bosons with equal masses.
If SUSY is to be  relevant for the physical world, it must be broken,
either softly or spontaneously.
That is, SUSY must broken because in experiment one does not observe
degenerate Bose-Fermi multiplets.
Thus the Higgs boson mass at the tree level receives quadratic corrections
that are limited by an incomplete cancelation of the SM particle and
its superpartner loops.

In the SM, only one Higgs doublet is required to give masses to the quarks
and leptons.
SUSY requires at least two Higgs doublets.
One is needed to give masses to the up-type quarks and the other to
the down-type quarks and leptons.
This is the minimal supersymmetric extension of the SM.
The minimal supersymmetric standard model (MSSM) is the most widely studied
supersymmetric extension of the SM.
The physical mass spectra of the Higgs bosons consist of two neutral
scalars, one neutral pseudoscalar, and a pair of charged scalars.
According to the tree level potential, the mass of the lightest scalar
Higgs boson have to be lighter than the Z boson mass, and the the charged
Higgs boson mass must be heavier than the W boson mass.

Recently radiative corrections to the Higgs boson masses have been
calculated by many authors.
The tree level constraints on the charged Higgs boson mass can be violated
when the top quark mass is heavy and the pseudoscalar Higgs boson mass is
light.
This means that radiative corrections to the charged Higgs boson mass can
be negative contributions.
However, these corrections to the tree level result are numerically very
small [1,2].
On the other hand radiative corrections to the lightest scalar Higgs boson
mass give a significant contribution and the mass of the lightest scalar
Higgs boson can be substantially heavier than the Z boson mass [3-5].

If one considers about more general extensions of the Higgs sector [6], the
tree level results in the MSSM can be changed.
The simplest extension is the next-to-minimal supersymmetric standard model
(NMSSM).
The Higgs sector in the NMSSM consists of two Higgs doublits $H_{1}$ and
$H_{2}$ and a Higgs singlet chiral superfield $N$.
In this model, the superpotential contains a new coupling among two Higgs
doublits and a Higgs singlet.
Thus the parameter $\mu$ in the superpontial of the MSSM can be generated
dynamically by the vacuum expectation value (VEV) of the singlet Higgs
field [7].
The NMSSM has ten real degrees of freedom.
After the Higgs mechanism takes place, three degrees of freedom correspond
to neutral ( $G$ ) and charged ( $G^{+}$, $G^{-}$ ) unphysical Goldstone
bosons, the other seven correspond to neutral
( $S_{1}$, $S_{2}$, $S_{3}$, $P_{1}$, $P_{2}$ ) and charged
( $C^{+}$, $C^{-}$ ) physical particles.
Here $S_{1}$, $S_{2}$, and $S_{3}$ are scalar Higgs bosons while
$P_{1}$ and $P_{2}$ are pseudoscalar Higgs bosons.
As a result, the effective number of parameters describing the Higgs
sector at the tree level is six.

As we have already mentioned above, the tree level result in the NMSSM
is a little different from that in the MSSM.
The mass of the charged Higgs boson may be lighter or heavier than the
W boson mass.
Also the upper bound on the lightest scalar Higgs boson mass can be heavier
than the Z boson mass [8].
Therefore the upper bound on the lightest scalar Higgs boson mass of the
NMSSM at the tree level is heavier than that of the MSSM.
Furthermore radiative corrections to the lightest scalar Higgs boson mass
lead to a important contributions.
It is well known that the largest contribution to the lightest scalar Higgs
boson mass comes from the top quark and scalar quark loops [9].
These effects are quite substantial when the top scalar quark mass is
much heavier than the top quark mass.
This implies that at least one top scalar quark mass is heavier than top
quark mass.
If the top quark and scalar quark masses are identical, the contribution
of radiative corrections to the lightest scalar Higgs boson mass vanishes.

Assuming the degeneracy of the left- and right-handed top scalar quarks,
radiative corrections to the lightest scalar Higgs boson mass are studied
in ref. [10,11].
Especially, the authers of ref. [11] performed a low energy renormalization
group analysis of the Higgs sector of the model, and they arrived at the
conclusion that the upper bound on the lightest scalar Higgs boson mass
is 123 GeV for $m_t$ = 180 GeV if the masses of the top scalar quark were
assumed to be degenerate at $M_{SUSY}$ = 1 TeV.

The main aim of this paper is to explore all the parameter space and
provide an accurate numerical evaluation for the upper bound on the
lightest scalar Higgs boson mass in the NMSSM.
We examine in detail the radiative corrections to the lightest scalar
Higgs boson mass using the 1-loop effective potential.
In general, the contribution of radiative corrections due to the bottom
quark and scalar quark loops to the tree level mass of the lightest
scalar Higgs boson is very small numerically.
Therefore we use the effective potential which includes the contribution
of the top quark and scalar quark loops assuming the nondegenerate
of the top scalar quark masses.
Thus the mass splitting among the top scalar quark give a
significant effect to the upper bound of the lightest scalar Higgs
boson mass.

In fact, there have been many studies about the upper bound on the lightest
scalar Higgs boson mass including radiative corrections in the NMSSM.
Unlike their calculations we especially use the top scalar quark mass
matrix containing the gauge terms in the 1-loop effective potential.
Therefore our formalism for the scalar Higgs boson mass matrix at the
1-loop level contains the contributions of the gauge sector.
After including radiative corrections, the upper bound on the lightest
scalar Higgs boson mass becomes dependent on $m_Q, m_T, A_t$, and $m_t$
as free parameters as well as the parameters at the tree level.
Of course our formalism for the mass matrix is given by the complicated
expression.
Nevertherless we believe that the upper bound on the lightest scalar Higgs
boson mass obtained by our formalism can provide a reliable result.

\section{HIGGS BOSON MASSES}

\hspace*{6.mm}
The scalar Higgs potential of the NMSSM comes from the auxiliary
F- and D-fields and the soft SUSY breaking terms.
The Higgs boson masses in the model at the tree level are given by
the relevant potential
\begin{equation}
	V_0 = V_F +V_D +V_{soft}
\end{equation}
which is expressed in terms of two Higgs doublet fields $H_1$ and
$H_2$ and a complex singlet field $N$ as
\begin{eqnarray}
	V_F &=& |\lambda|^2[(|H_1|^2+|H_2|^2)|N|^2+|H_1 H_2|^2]
		+ |k|^2|N|^4 \cr
   		& &-(\lambda k^*H_1H_2N^{*2}+H.c.) \ , \cr
        V_D &=& {g_2^2\over 8}
		(H_2^\dagger\hat\sigma H_2+ H_1^\dagger\hat\sigma H_1)^2
                +{g_1^2\over 8}(|H_2|^2-|H_1|^2)^2 \ , \\
	V_{soft} &=& m_{H_1}^2|H_1|^2+m_{H_2}^2|H_2|^2+m_N^2|N|^2
		-(\lambda A_\lambda H_1 H_2 N+H.c.) \cr
   		& &\mbox{} - ({1\over 3} kA_kN^3+H.c.) \ . \nonumber
\end{eqnarray}
In the above expressions, $g_1$ and $g_2$ are the U(1) and SU(2) gauge
coupling constants, respectively,
$\hat\sigma = (\sigma^1,\sigma^2,\sigma^3)$
are the Pauli matrices, $A_k$ and $A_\lambda$ are the trilinear soft SUSY
breaking parameters, and $H_1 H_2 = H_1^0 H_2^0-H^-H^+$.
The parameters $\lambda$ and $k$ come from the trilinear couplings in
the part of the superpotential of the NMSSM
\[
        \lambda N H_1 H_2-{k\over 3} N^3  \ .
\]

We assume that only neutral components of the three Higgs fields
$H_1$, $H_2$, and $N$ acquire vacuum expectation values
$v_1$, $v_2$, and $x$, respectively. Without loss of generality, we also
assume that the vacuum expectation values are real and positive.
The conditions that the Higgs potential becomes minimum at $H_1 = v_1$,
$H_2 = v_2$, and $N = x$ yield three constraints, which can eliminate
the soft SUSY breaking parameters $m_{H_1}$, $m_{H_2}$, and $m_N$ from
the potential $V_0$.

Then the mass spectra of the Higgs bosons can be expressed in terms of
six parameters, $\lambda$, $k$, $A_\lambda$, $A_k$, $x$,
and $\tan \beta = v_2 / v_1$. Note that $m_Z^2 = (g_1^2+g_2^2)v^2/2$ for
$v = \sqrt{v_1^2 + v_2^2}$ = 175 GeV.
Collecting the real parts of the neutral Higgs fields from $V_0$, one
can obtain the mass spectrum of the scalar Higgs bosons at the tree level.
It is given by a symmetric mass matrix, whose elements are as follows:
\begin{eqnarray}
      M^0_{11} & = & (m_Z \cos \beta)^2
            + \lambda x \tan \beta (A_{\lambda} + kx)  \ , \cr
      M^0_{22} & = & (m_Z \sin \beta)^2
            + \lambda x \cot \beta (A_{\lambda} + kx)  \ , \cr
      M^0_{33} & = & (2 k x)^2 - k x A_k
            + {\lambda \over 2x} v^2 A_{\lambda} \sin 2 \beta
                        \ , \\
      M^0_{12} & = & (\lambda^2 v^2 - {1 \over 2} m_Z^2) \sin 2 \beta
            - \lambda x (A_{\lambda} +kx)  \ , \cr
      M^0_{23} & = & 2 \lambda^2 x v \sin \beta
            - \lambda v \cos \beta (A_{\lambda} + 2 k x) \ , \cr
      M^0_{13} & = & 2 \lambda^2 x v \cos \beta
            - \lambda v \sin \beta (A_{\lambda} +2kx) \ . \nonumber
\end{eqnarray}
When the above mass matrix is diagonalized, we would obtain squared
masses for the three scalar Higgs bosons.
The smallest eigenvalue yields the mass of the lightest scalar
Higgs boson, $m_{S_1}$.

The tree-level masses obtained from $V_0$ are subject to radiative
corrections. These corrections are contributed in supersymmetric models
by the quark and scalar quark loops.
According to the Coleman-Weinberg mechanism [12], the 1-loop corrections
are given by the effective potential containing the mass squared
matrix ${\cal M}^2$ of the scalar quark,
\begin{equation}
	V_1 = \frac{1}{64\pi^2} Str {\cal M}^4 \left(
           \log \frac{{\cal M}^2}{\Lambda^2} - \frac{3}{2} \right) \ ,
\end{equation}
where the supertrace is defined as
\begin{equation}
	Str f({\cal M}^2) = \sum_i (-1)^{2J_i} (2J_i+1) f(m_i^2) \ .
\end{equation}
The squared masses of the quark and scalar quark are denoted by $m^2_i$
and their spins are $J_i = 0$ for the quark and $J_i = 1/2$ for the
scalar quark, respectively.
The arbitrary scale $\Lambda$ is taken to be $M_{SUSY}$ = 1 TeV.
Thus the expression for the 1-loop effective potential which corresponds
to taking into account the radiative corrections due to the top quark
and scalar quark loops is given as
\begin{eqnarray}
   V_1 & = &\frac{3}{32\pi^2} m_{\tilde{q_1}}^4
   \left (\log {m_{\tilde{q_1}}^2 \over \Lambda^2} - {3\over 2} \right )
              + \frac{3}{32\pi^2} m_{\tilde{q_2}}^4
\left(\log {m_{\tilde{q_2}}^2 \over \Lambda^2}-{3\over 2} \right)
       \nonumber  \\
            &   &\mbox{} - \frac{3}{16\pi^2} m_q^4
        \left (\log {m_q^2 \over \Lambda^2}  - {3\over 2} \right )  \ ,
\end{eqnarray}
where $m_{\tilde{q_1}}$ and $m_{\tilde{q_2}}$ are the masses of the
top scalar quark, $m_{\tilde{q_1}} < m_{\tilde{q_2}}$
and $m_t$ is the top quark mass.

In terms of $V_1$, the mass matrix of the scalar Higgs bosons including
radiative corrections at the 1-loop level is given by
\begin{equation}
	M^1_{ij} = \left. \left(
	{\partial^2 V_1 \over \partial \phi_i \partial \phi_j}
	- {1\over v_i} {\partial V_1 \over \partial \phi_j}
	\delta_{ij} \right)
	\right| _{H^0_1 = v_1,\, H^0_2 = v_2,\, N = x} \ ,
\end{equation}
where $\phi_i$ are the conventional notations for the real and
imaginary parts of the Higgs fields.
The first derivative term of the effective potential is minimization
condition.

In order to calculate the radiative corrections, one has to know the
squared masses of the scalar quarks in $V_1$.
They are given as the eigenvalues of a $4 \times 4$ Hermitian matrix
${\cal M}^2$ for the third generation.
The scalar quark mass matrix may be broken up into four $2 \times 2$
submatrices, because only two block-diagonal submatrices depend
on the neutral Higgs sector and thus are nonzero.
The two eigenvalues of the upper-left block-diagonal submatrix are the
squared masses of the left- and right-handed top scalar quarks,
and those of the lower-right block-diagonal submatrix the squared masses
of the left- and right-handed bottom scalar quarks.
Although the contribution of the bottom quark and scalar quark loops for
large $\tan \beta$ is not negligible, generally the leading radiative
corrections to the scalar Higgs bosons are the contribution of the top
quark and scalar quark loops.

Taking only the top quark and scalar quark into account, the matrix
elements of ${\cal M}^2$ depending on the neutral Higgs sector are:
\begin{eqnarray}
        {\cal M}_{11} &=& m_Q^2 + h_t^2 |H_2^0|^2
        - {g_1^2 \over 12} ( |H_1^0|^2 - |H_2^0|^2)
        + {g_2^2 \over 4} (|H_1^0|^2 - |H_2^0|^2)  \ , \cr
        & & \cr
        {\cal M}_{12} &=& h_t (\lambda N H_1^0 + A_t H_2^{0*})  \ , \cr
        & & \cr
        {\cal M}_{22} &=& m_T^2 + h_t^2 |H_2^0|^2
        + {g_1^2 \over 3}(|H_1^0|^2 - |H_2^0|^2)  \ ,
\end{eqnarray}
where $h_t$ is the Yukawa coupling of top quark,
in the part of the superpotential of the NMSSM
\[
        h_t Q t^c H_2  \ ,
\]
and $m_Q$ and $m_T$ are the soft SUSY breaking scalar quark masses and
$t^c$ is the charge conjugate of the right-handed top quark.
We assume that the soft SUSY breaking scalar quark masses are real and
positive.

The eigenvalues of the matrix are expressed by the top scalar quark masses
\begin{eqnarray}
        m_{\tilde{t}_{1,2}}^2 & = & m_t^2 + {1\over 2}(m_Q^2 + m_T^2)
        + {1\over 4} m_Z^2 \cos 2 \beta            \cr
        &   & \mbox{} \pm
        \left [\left ({1\over 2}(m_Q^2 -m_T^2)
        +({2\over 3} m_W^2 - {5\over 12} m_Z^2) \cos 2 \beta \right )^2
        + m_t^2(A_t + \lambda x \cot \beta)^2 \right]^{{1\over 2}}
\end{eqnarray}
where $m_t$ = $h_t v \sin \beta$ and $m_W^2 = g_2^2(v_1^2 + v_2^2)/2$.
From the formula we find that the masses of the top scalar quark are
symmetric under interchange $m_Q \leftrightarrow m_T$.

Substituting these expressions into $V_1$, one can now calculate the
radiative corrections.
After lengthy calculation, the elements of the mass matrix for the scalar
Higgs bosons including the 1-loop corrections are obtained as
\begin{eqnarray}
        M_{11}^1 &=& {3 \over 8 \pi^2} \Delta_1^2
        g(m_{\tilde{t_1}}^2,m_{\tilde{t_2}}^2)
        +{3 m_Z^4 \cos^2 \beta \over 128 \pi^2 v^2}
        \log {m_{\tilde{t_1}}^2 m_{\tilde{t_2}}^2\over \Lambda^4}  \cr
        & &\mbox{} + {3 \over 16 \pi^2 v^2}
        \left [ {2 m_t^2 A_t \lambda x \over \sin 2 \beta}
        - ({4 \over 3} m_W^2 - {5 \over 6} m_Z^2)^2 \cos^2 \beta \right ]
        f(m_{\tilde{t_1}}^2,m_{\tilde{t_2}}^2)       \cr
        & &\mbox{} + {3 \over 16 \pi^2 v} m_Z^2 \cos \beta
  \left ({\Delta_1 \over m_{\tilde{t_1}}^2 - m_{\tilde{t_2}}^2} \right )
        \log {m_{\tilde{t_1}}^2 \over m_{\tilde{t_2}}^2}  \cr
        & &  \cr
        M_{22}^1 &=& {3 \over 8 \pi^2} \Delta_2^2
        g(m_{\tilde{t_1}}^2,m_{\tilde{t_2}}^2)
        - {3 m_t^4 \over 4 \pi^2 v^2 \sin^2 \beta}
        \log {m_t^2 \over \Lambda^2}  \cr
        & &\mbox{} + {3 \over 16 \pi^2 v^2}
        \left [ {m_t^2 A_t \lambda x \cot \beta \over \sin^2 \beta}
        - ({4 \over 3} m_W^2 - {5 \over 6} m_Z^2)^2 \sin^2 \beta \right ]
        f(m_{\tilde{t_1}}^2,m_{\tilde{t_2}}^2)       \cr
        & &\mbox{} + {3 \over 16 \pi^2 v}
        \left ({4 m_t^2 \over \sin \beta} - m_Z^2 \sin \beta \right )
   \left ({\Delta_2 \over m_{\tilde{t_1}}^2 - m_{\tilde{t_2}}^2} \right )
        \log {m_{\tilde{t_1}}^2 \over m_{\tilde{t_2}}^2}  \cr
        & &\mbox{} + {3 \over 32 \pi^2 v^2}
        \left ({2 m_t^2 \over \sin \beta}
        - {1 \over 2} m_Z^2 \sin \beta \right )^2
        \log {m_{\tilde{t_1}}^2 m_{\tilde{t_2}}^2 \over \Lambda^4}  \cr
        & &  \cr
        M_{33}^1 &=& {3 \over 8 \pi^2} m_t^4 \lambda^2 \cot^2 \beta
        (A_t + \lambda x \cot \beta)^2
        g(m_{\tilde{t_1}}^2,m_{\tilde{t_2}}^2)        \cr
        & &\mbox{} + {3 \over 16 \pi^2 x} m_t^2 A_t \lambda \cot \beta
        f(m_{\tilde{t_1}}^2,m_{\tilde{t_2}}^2)       \cr
        & & \cr
        M_{12}^1 &=& {3 \over 8 \pi^2} \Delta_1 \Delta_2
        g(m_{\tilde{t_1}}^2,m_{\tilde{t_2}}^2)
        + {3 m_Z^2 \sin 2 \beta \over 256 \pi^2 v^2}
        \left( {4 m_t^2 \over \sin^2 \beta } -m_Z^2 \right)
        \log {m_{\tilde{t_1}}^2 m_{\tilde{t_2}}^2 \over \Lambda^4}  \cr
        & &\mbox{} + {3 \over 32 \pi^2 v^2}
        \left [({4 \over 3} m_W^2 - {5 \over 6} m_Z^2)^2 \sin 2 \beta
        - {2 m_t^2 A_t \lambda x \over \sin^2 \beta} \right]
        f(m_{\tilde{t_1}}^2,m_{\tilde{t_2}}^2)       \cr
        & &\mbox{} + {3 \over 32 \pi^2 v} \left [m_Z^2 \cos \beta \Delta_2
     + ({4 m_t^2 \over \sin^2 \beta} - m_Z^2) \sin \beta \Delta_1 \right]
        {1 \over (m_{\tilde{t_1}}^2 - m_{\tilde{t_2}}^2)}
        \log {m_{\tilde{t_1}}^2 \over m_{\tilde{t_2}}^2}  \cr
        & & \cr
        M_{23}^1 &=& {3 \over 8 \pi^2} \Delta_2 m_t^2 \lambda \cot \beta
        (A_t + \lambda x \cot \beta)
        g(m_{\tilde{t_1}}^2,m_{\tilde{t_2}}^2) \cr
        & &\mbox{} + {3 m_t^2 \lambda \cot \beta \over 32 \pi^2}
        \left ({4 m_t^2 \over \sin \beta} - m_Z^2 \sin \beta \right)
        \left ({ A_t + \lambda x \cot \beta
        \over m_{\tilde{t_1}}^2 - m_{\tilde{t_2}}^2} \right)
        \log {m_{\tilde{t_1}}^2 \over m_{\tilde{t_2}}^2}  \cr
        & &\mbox{}
        - {3 m_t^2 A_t \lambda \cot \beta \over 16 \pi^2 v \sin \beta}
        f(m_{\tilde{t_1}}^2,m_{\tilde{t_2}}^2)       \cr
        & &  \cr
        M_{13}^1 &=& {3 \over 8 \pi^2} \Delta_1 m_t^2 \lambda \cot \beta
        (A_t + \lambda x \cot \beta)
        g(m_{\tilde{t_1}}^2,m_{\tilde{t_2}}^2) \cr
        & &\mbox{}
        + {3 m_Z^2 m_t^2 \lambda \cos \beta \cot \beta \over 32 \pi^2 v}
        \left ({ A_t + \lambda x \cot \beta
        \over m_{\tilde{t_1}}^2 - m_{\tilde{t_2}}^2} \right)
        \log {m_{\tilde{t_1}}^2 \over m_{\tilde{t_2}}^2}  \cr
        & &\mbox{}
        - {3 m_t^2 \lambda \over 16 \pi^2 v \sin \beta}
        (A_t + 2 \lambda x \cot \beta)
        f(m_{\tilde{t_1}}^2,m_{\tilde{t_2}}^2)
\end{eqnarray}
with
\begin{eqnarray}
\Delta_1  &=& {m_t^2 \lambda x \over v \sin \beta}
              (A_t + \lambda x \cot \beta)             \cr
          & &\mbox{} + {\cos \beta \over 2 v}
          \left [ (m_Q^2 - m_T^2) + ({4\over 3} m_W^2
          - {5\over 6} m_Z^2) \cos 2 \beta \right]
          ({4\over 3} m_W^2 - {5\over 6} m_Z^2)  \ , \cr
\Delta_2  &=& {m_t^2 A_t\over v \sin \beta}
              (A_t + \lambda x \cot \beta)             \cr
          & &\mbox{} - {\sin \beta \over 2 v}
          \left [ (m_Q^2 - m_T^2) + ({4\over 3} m_W^2
          - {5\over 6} m_Z^2) \cos 2 \beta \right]
          ({4\over 3} m_W^2 - {5\over 6} m_Z^2)  \ .
\end{eqnarray}
The two functions $f$ and $g$ are defined as
\begin{eqnarray}
        f(m_1^2,m_2^2) &=& {1 \over (m_2^2-m_1^2)}
\left[  m_1^2 \log {m_1^2 \over \Lambda^2} -m_2^2
\log {m_2^2 \over \Lambda^2}
        \right] + 1  \ , \cr
        g(m_1^2,m_2^2) &=& {1 \over (m^2_1 - m^2_2)^3}
        \left [(m_1^2+m_2^2) \log {m_2^2 \over m_1^2}
        - 2 (m_2^2 - m_1^2) \right ] \ .
\end{eqnarray}

Note that these calculations have been done by several authors, either
with some approximations or without [13-16].
Nevertheless, their formulas for the mass matrix of the scalar Higgs boson
have not contain the contributions of the gauge sector at the 1-loop level.
Unlike their calculation we have derived a formula for the mass matrix of
the scalar Higgs boson including the contribution of the gauge sector.
Furthermore, the number of the parameter for the mass matrix at the 1-loop
level is the same that of their formula.
When radiative corrections to the tree level mass of the scalar Higgs
boson are included, the values of the top quark Yukawa coupling are
determined by the equation, $m_t = h_t v \sin \beta$.
As a consequence, one could have taken the top quark mass instead of the
top quark Yukawa coupling as a parameter.
Thus the mass matrix at the 1-loop level contains $m_Q, m_T, A_t$, and
$m_t$ as free parameters as well as the parameters at the tree level.

\section{NUMERICAL ANALYSIS}

\hspace*{6.mm}
Now let us consider the mass splitting among the top scalar quark.
After considering all conditions and assumption in the NMSSM,
we initiate a numerical study for the mass spliting among the top scalar
quark.
In the renormalization group analysis of the NMSSM, the maximum values of
the dimensionless parameters $\lambda$ and $k$ are 0.87 and 0.63,
respectively.
From the experimental result of the CDF and D0 collaborations, the range
of top quark mass is given as $176 \pm 8 \pm 10$ and
$199^{+19}_{-21} \pm 22$ GeV, respectively [17].
Therefore, we take the top quark mass to be 175 GeV.
The range for $\tan \beta$ is defined in terms of many ways.
It is demonstrated that the renormalization group analysis of radiative
symmetry breaking always leads to $\tan \beta$ values that are
larger than 1 [8].
In our argumentation, we consider the ratio of VEV's $\tan \beta$ =
$v_2/v_1$ to be in the range
\begin{equation}
       2 \le \tan\beta \le {m_t \over m_b} \ .
\end{equation}
Thus the maximum value of the free parameter $\tan \beta$ is about 39 for
$m_t$ = 175 and $m_b$ = 4.5 GeV.
From the values of $\tan \beta$ and $m_t$, the value of the coupling
$\lambda$ is given by the analysis of the renormalization group equation.

Here we assume that $m_{\tilde{t_1}}$ is heavier than the mass of the top
quark.
For the convenience we then define the ratio of the top scalar quark
masses by R,
\begin{equation}
       0 < {\rm R} = {m_{\tilde{t_1}} \over m_{\tilde{t_2}}} < 1 \ .
\end{equation}
Thus R is the inverse of the mass splitting among the top scalar quark.
This implies that the mass splitting among the top scalar quark increases
with decreasing R values, vice versa.
The curves for the maximum and minimum of R are denotted by
${\rm R}_{\rm maximum}$ and ${\rm R}_{\rm minimum}$ in Fig. 1.
As can be seen in the figure, these curves decreases monotonically as
the value of $A_t$ increases to 3000 GeV.
That is, the mass splitting among the top scalar quark always increases
with increasing $A_t$ value.

Next we now wish to evaluate the numerical result for the upper
bound on the lightest scalar Higgs boson mass including radiative
corrections in the NMSSM.
The radiatively corrected squared masses of the scalar Higgs bosons
are obtained as the eigenvalues of $M^0 + M^1$.
It is not easy to obtain an analytic expression for the
smallest eigenvalue, $m^2_{S_1}$, in terms of the various parameters
that are present in $M^0 + M^1$.
The lightest scalar Higgs boson mass at the 1-loop level is obtained by
the numerical diagonalizion of the mass matrix of the scalar Higgs boson.

Fig. 2 shows the upper bounds on the lightest scalar Higgs boson mass
including the radiative corrections due to the top quark and scalar
quark loops, as a function of $x$, for $m_t$ = 175 GeV and
$\tan \beta = 2$, $0 < \lambda \le 0.87$, $0 < k \le 0.63$, and
100 GeV $\le A_\lambda (A_k, m_Q, m_T) \le$ 1000 GeV.
The bounds on $m_{S_1}$ for $A_t$ = 1000, 2000, and 3000 GeV have
maximum values as the value of $x$ approach 1000 GeV.
The bound for $A_t$ = 2000 GeV is greater than those for the other values
of $A_t$ when the value of $x$ is given by 1000 GeV.
We can infer that the bound on $m_{S_1}$ for $\tan \beta$ = 2 and $x$ = 1000 GeV
is not increase thought the value of $A_t$ is greater than 3000 GeV.
Therefore the upper bound on $m_{S_1}$ for $\tan \beta$ = 2 and $x$ = 1000
GeV is determined in terms of a value between 1000 and 3000 GeV.
Thus we come to the conclusion that the upper bound on $m_{S_1}$ for
$\tan \beta$ = 2 and $x$ = 1000 GeV can be greater than 143 GeV
in an energy between $A_t$ = 1000 and 3000 GeV.
In Fig. 3 we plot the upper bound on $m_{S_1}$ for $\tan \beta$ = 6 when
the other parameters are given by the same parameter space as those of
Fig. 2.
We find from Fig. 3 that the upper bound on $m_{S_1}$ for $x$ = 300 GeV and
$A_t$ = 2000 GeV is about 151 GeV.
The value of $m_{S_1}$ = 151 GeV is greater than that for
$\tan \beta$ = 2 in Fig. 2.
Therefore, let us calculate the upper bound on $m_{S_1}$ for
$\tan \beta$ = 10 through the same parameter space as those in Fig. 3.
Fig. 4 shows that the upper bound on $m_{S_1}$ for $x$ = 160 GeV and
$A_t$ = 2000 GeV is about 148 GeV.
But the value of $m_{S_1}$ = 148 GeV is smaller than that for
$\tan \beta$ = 6 in Fig. 3.
From Fig. 2, Fig. 3, and Fig. 4, we surmise that the upper bound on
$m_{S_1}$ for $\tan \beta >$ 10 is smaller than that on $m_{S_1}$
for $\tan \beta$ = 6.

In Fig. 5  we plot the upper bound on $m_{S_1}$ at the 1-loop level,
as a function of $A_t$, for $m_t$ = 175 GeV, $0 < \lambda \le 0.87$,
$0 < k \le 0.63$, and 100 GeV $\le A_\lambda (A_k, m_Q, m_T, x) \le$
1000 GeV.
The upper bound on $m_{S_1}$ for $\tan \beta$ = 6 is not increase though
the value of $A_t$ is greater than 3000 GeV.
Again we conject that $m_{S_1}$ have a maximum value in a value between
$\tan \beta$ = 2 and 10.
Fig. 6 shows that the upper bounds on $m_{S_1}$ at the tree level and
1-loop level, as a function of $\tan \beta$, for $m_t$ = 175 GeV,
$0 < \lambda \le 0.87$, $0 < k \le 0.63$,
100 GeV $\le A_\lambda (A_k, m_Q, m_T, x) \le$ 1000 GeV,
and 0 GeV $\le A_t \le$ 3000 GeV.
The absolute upper bound on $m_{S_1}$ at the tree level and 1-loop level
are 101 and 156 GeV for $\tan \beta \approx$ 2.7, respectively.

\section{RESULTS AND CONCLUSIONS}

\hspace*{6.mm}
In order to devise effective search strategies for the detection of
Higgs particle, the study for the lightest scalar Higgs boson mass is very
important.
Particularly, supersymmetric models impose strong constraints on the
lightest Higgs boson mass.
In the NMSSM the upper bound on $m_{S_1}$ at the tree level is about 130
GeV for all the parameter space.
By the renormalization group analysis the improved upper bound on
$m_{S_1}$ is about 101 GeV for $m_t \approx$ 175 GeV.

We know from the previous papers that the top quark and scalar quark loops
have a positive effect on the lightest Higgs boson mass when radiative
corrections to the Higgs sector of the tree level are considered.
But authors of the papers calculated the lightest scalar Higgs boson mass
with some approximations.
Here we examine in detail radiative corrections to the lightest Higgs boson
mass without some approximations.

The input parameters at the 1-loop level as well as at the tree
level are varied independently.
We find from Fig.1 that $A_t$ play an important role in the mass splitting
among the top scalar quark.
Here the value of $A_t$ is allowed to vary from 0 GeV to 3000 GeV.
The free parameters, $A_\lambda$, $A_k$, $m_Q$, $m_T$, and $x$ are limited
as masses below 1000 GeV.
In this case, the upper limits on the top scalar quark masses for
$m_t$ = 175 GeV are approximately given by $\sim$ 1000 GeV.
We are investigated the dependence of the lightest scalar Higgs boson mass
on all input parameters.
We calculated radiative corrections to the lightest Higgs boson mass using
the 1-loop effective potential in the NMSSM.
Here we consider the contribution of the top quark and scalar quark loops
for the effective potential.

We take the mass matrix of the top scalar quark including the gauge term.
Furthermore the mass matrix of the top scalar quark are given as
the nongenerate state.
Additionally we assume that the lower limits on the top scalar quark
masses are greater than the top quark mass.
Thus radiative correctons to the lightest Higgs boson mass contain the
effect of the mass spliting among the top scalar quark and the
contribution of the gauge sector.
An analytic formula for the scalar Higgs boson mass matrix including these
effects are derived.
The upper bound on $m_{S_1}$ containing these effects is calculated
by our formula.

We find from Fig. 5 that the upper bound on $m_{S_1}$ have a maximum value as
$A_t$ approach a particular value.
But the value of $A_t$ can not exceed 3000 GeV when the upper bound on
$m_{S_1}$ is maximised.
In other to find the absolute upper bound as a function of $\tan \beta$
including radiative corrections, we maximized over the parameter space
using numerical analysis.
The result of our calculation are presented in Fig. 6.
We also find from Fig. 6 that there exists the absolute upper bound
on $m_{S_1}$ of about 156 GeV.

\vskip 0.3 in
\noindent
{\large {\bf ACKNOWLEDGEMENTS}}

This work is supported in part by the Basic Science Research Institute Program,
Ministry of Education, BSRI-96-2442.
\vskip 0.2 in


\vfil\eject
\noindent
{\bf Figure Captions}
\vskip 0.3 in
\noindent
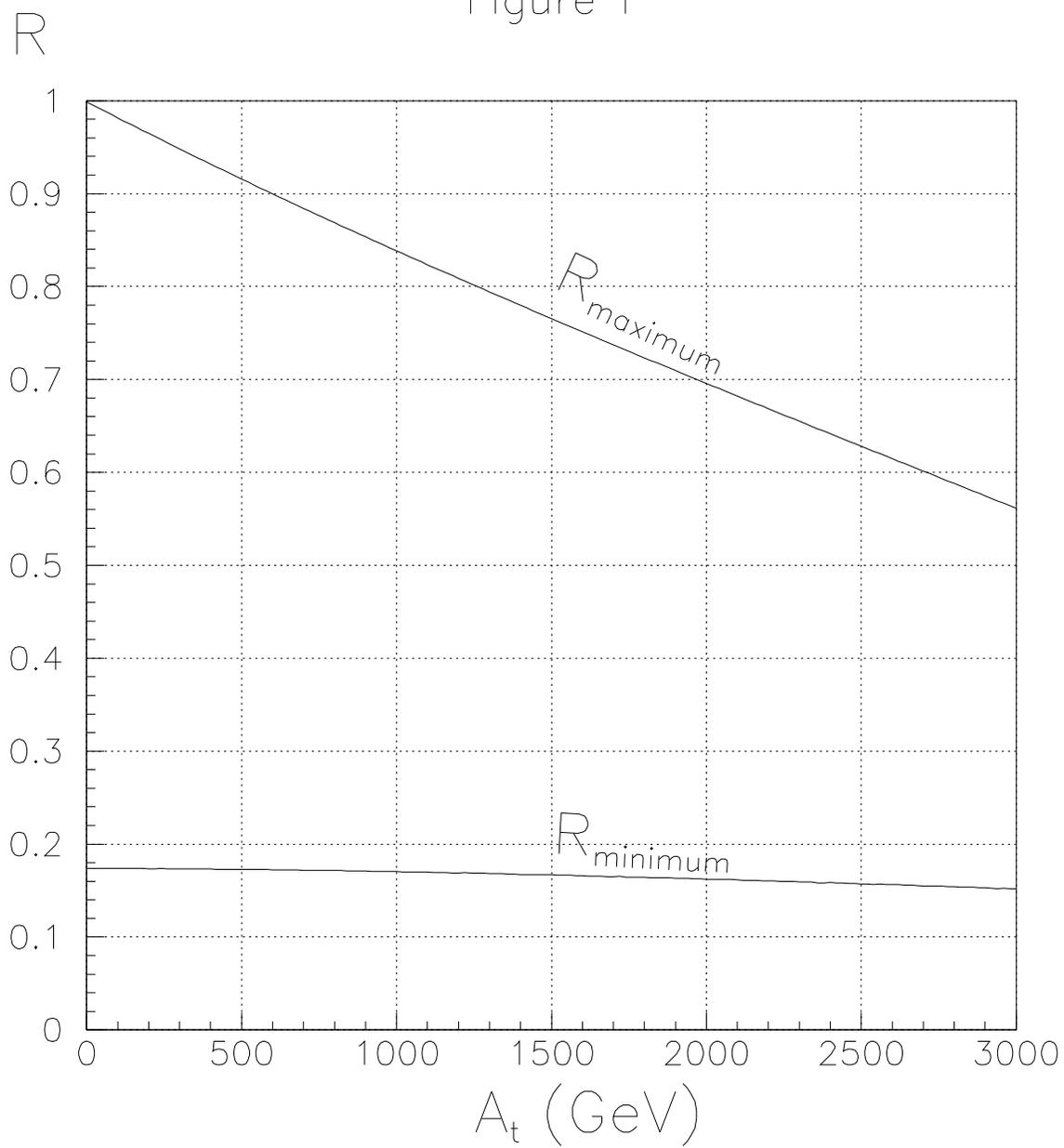
Fig. 1. \  The mass ratio of the top scalar quarks, as a function
of $A_t$, for $m_t$ = 175 GeV, $0 < \lambda \le 0.87$,
$2 \le \tan \beta \le 10$, and 100 GeV $\le m_Q (m_T, x) \le$ 1000 GeV.
The upper and lower curves are the maximum and minimum of R, respectively.
\vskip 0.2 in
\noindent
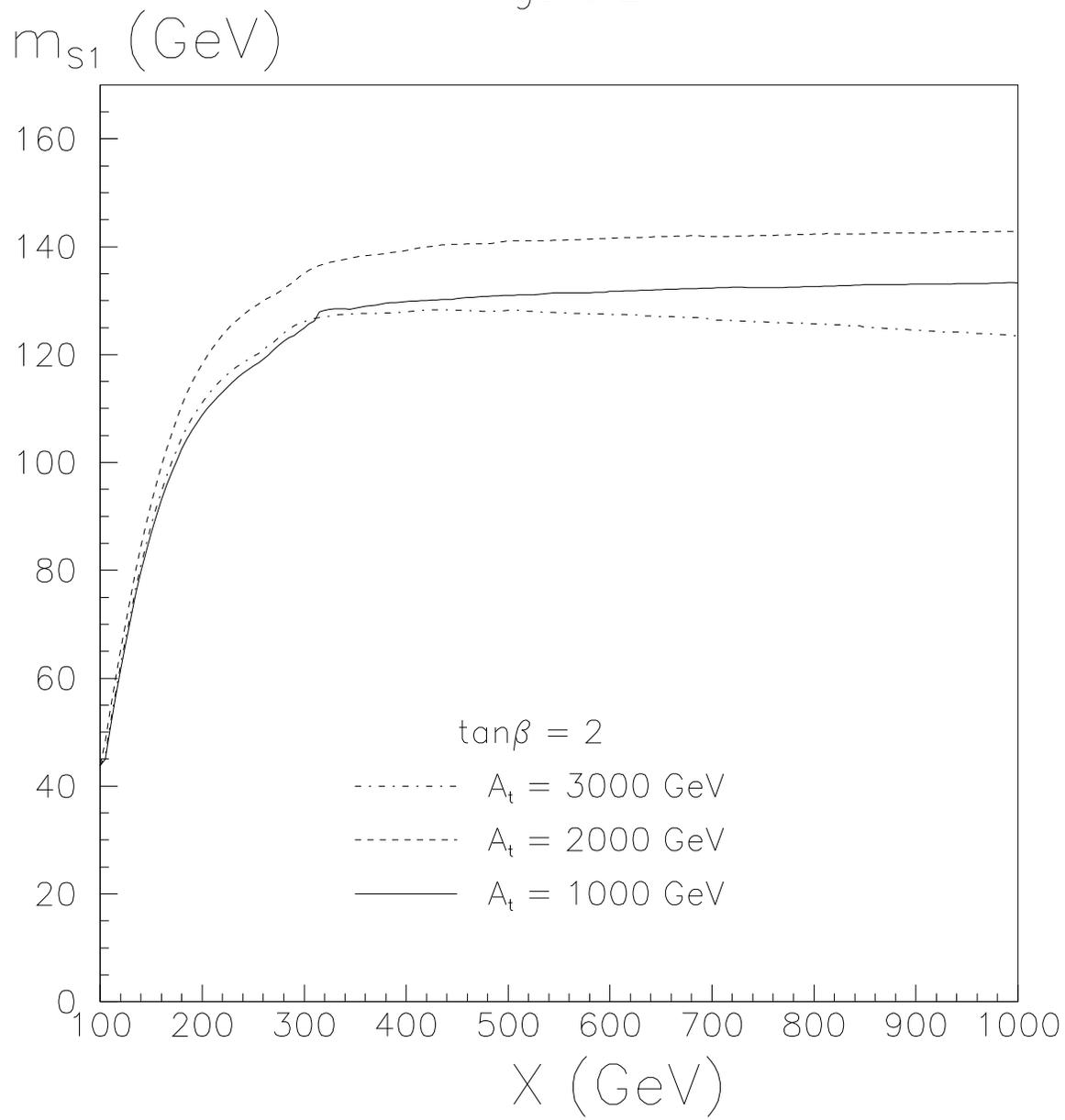
Fig. 2. \  The upper bound on the lightest Higgs boson mass at the 1-loop
level, as a function of $x$, for $m_t$ = 175 GeV, $\tan \beta = 2$,
$0 < \lambda \le 0.87$, $0 < k \le 0.63$, and
100 GeV $\le A_\lambda (A_k, m_Q, m_T) \le$ 1000 GeV.
These curves correspond to $A_t$ = 1000, 2000, and 3000 GeV, respectively.
\vskip 0.2 in
\noindent
Fig. 3. \  The same as Fig. 2, except for $\tan \beta$ = 6.
\vskip 0.2 in
\noindent
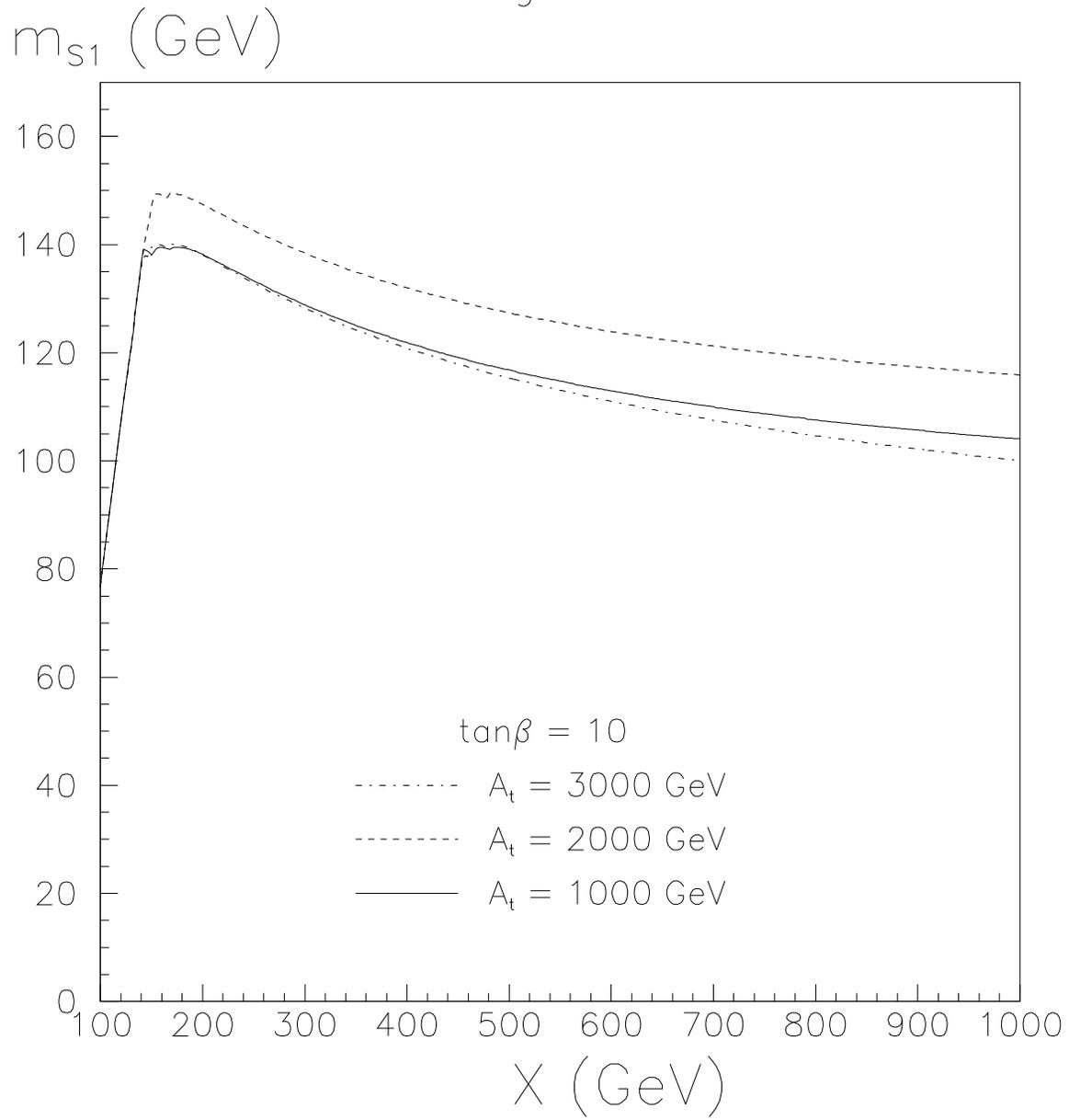
Fig. 4. \  The same as Fig. 2, except for $\tan \beta$ = 10.
\vskip 0.2 in
\noindent
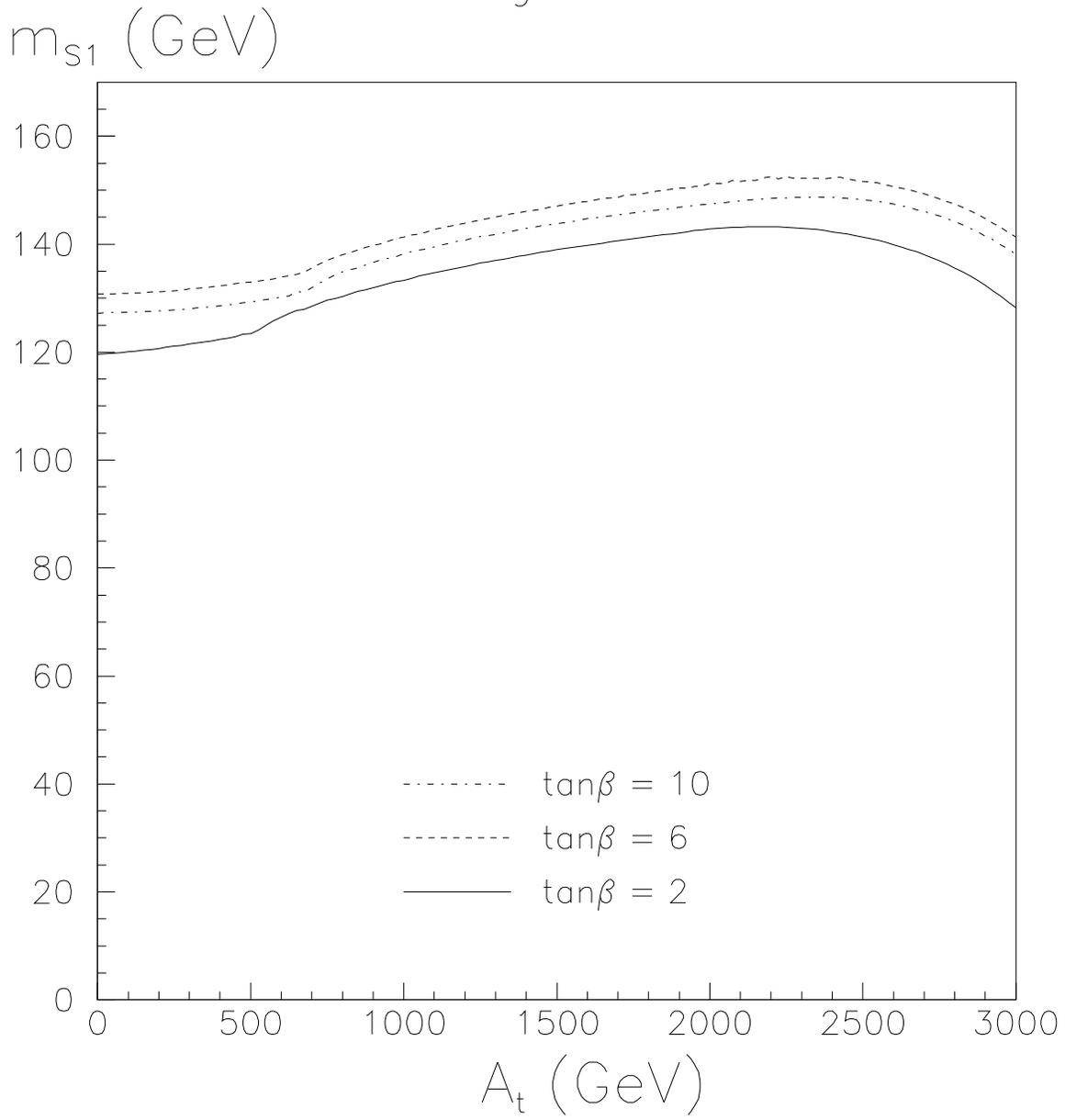
Fig. 5. \  The upper bound on the lightest Higgs boson mass at the 1-loop
level, as a function of $A_t$, for $m_t$ = 175 GeV,
$0 < \lambda \le 0.87$, $0 < k \le 0.63$, and
100 GeV $\le A_\lambda (A_k, m_Q, m_T, x) \le$ 1000 GeV.
These curves correspond to $\tan \beta$ = 2, 6, and 10, respectively.
\vskip 0.2 in
\noindent
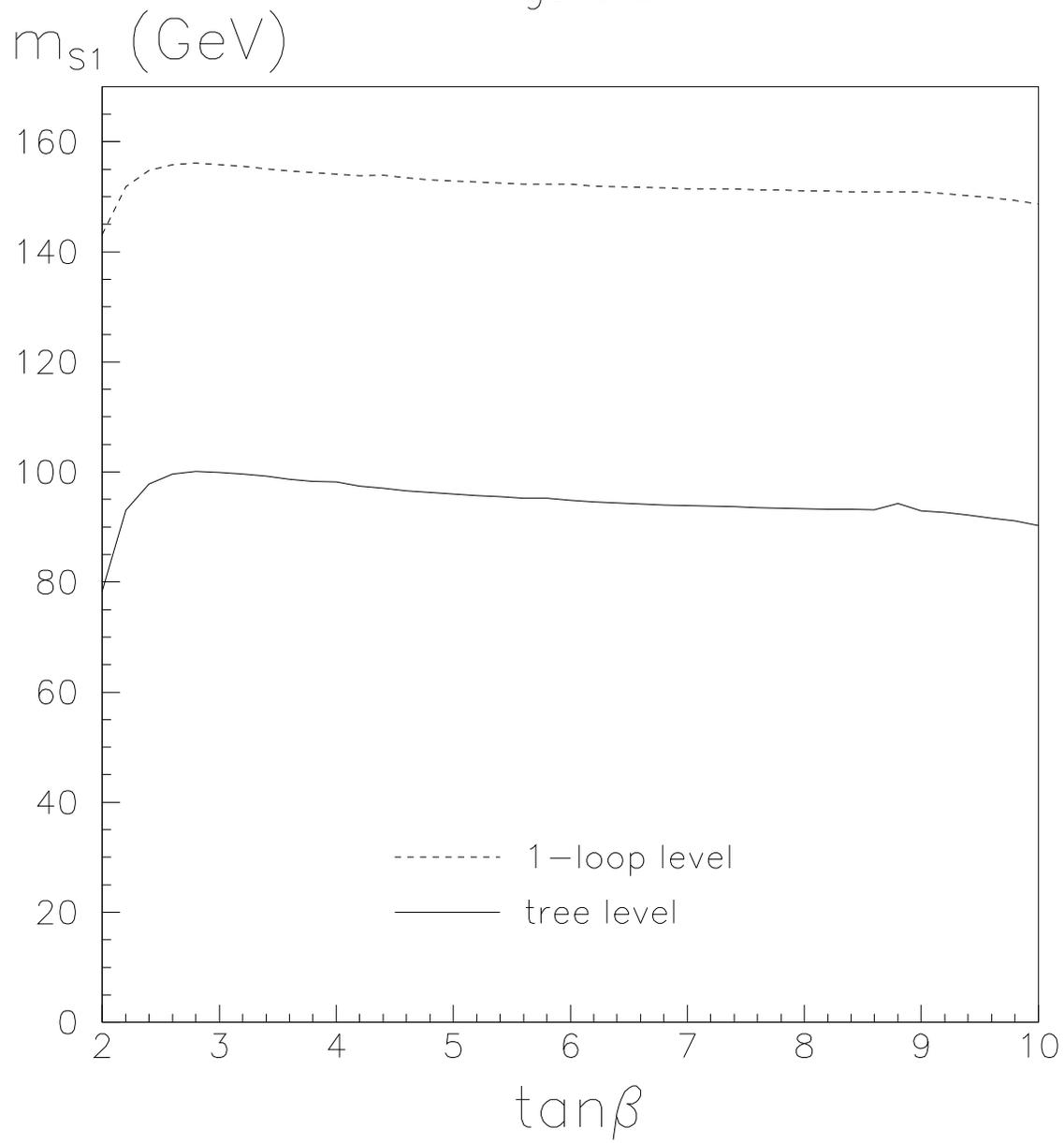
Fig. 6. \  The absolute upper bound on the lightest Higgs boson mass,
as a function of $\tan \beta$, for $m_t$ = 175 GeV,
$0 < \lambda \le 0.87$, $0 < k \le 0.63$,
100 GeV $\le A_\lambda (A_k, m_Q, m_T, x) \le$ 1000 GeV,
and 0 GeV $\le A_t \le$ 3000 GeV.
These curves correspond to the tree level and 1-loop level, respectively.
\vskip 0.2 in
\vfil\eject
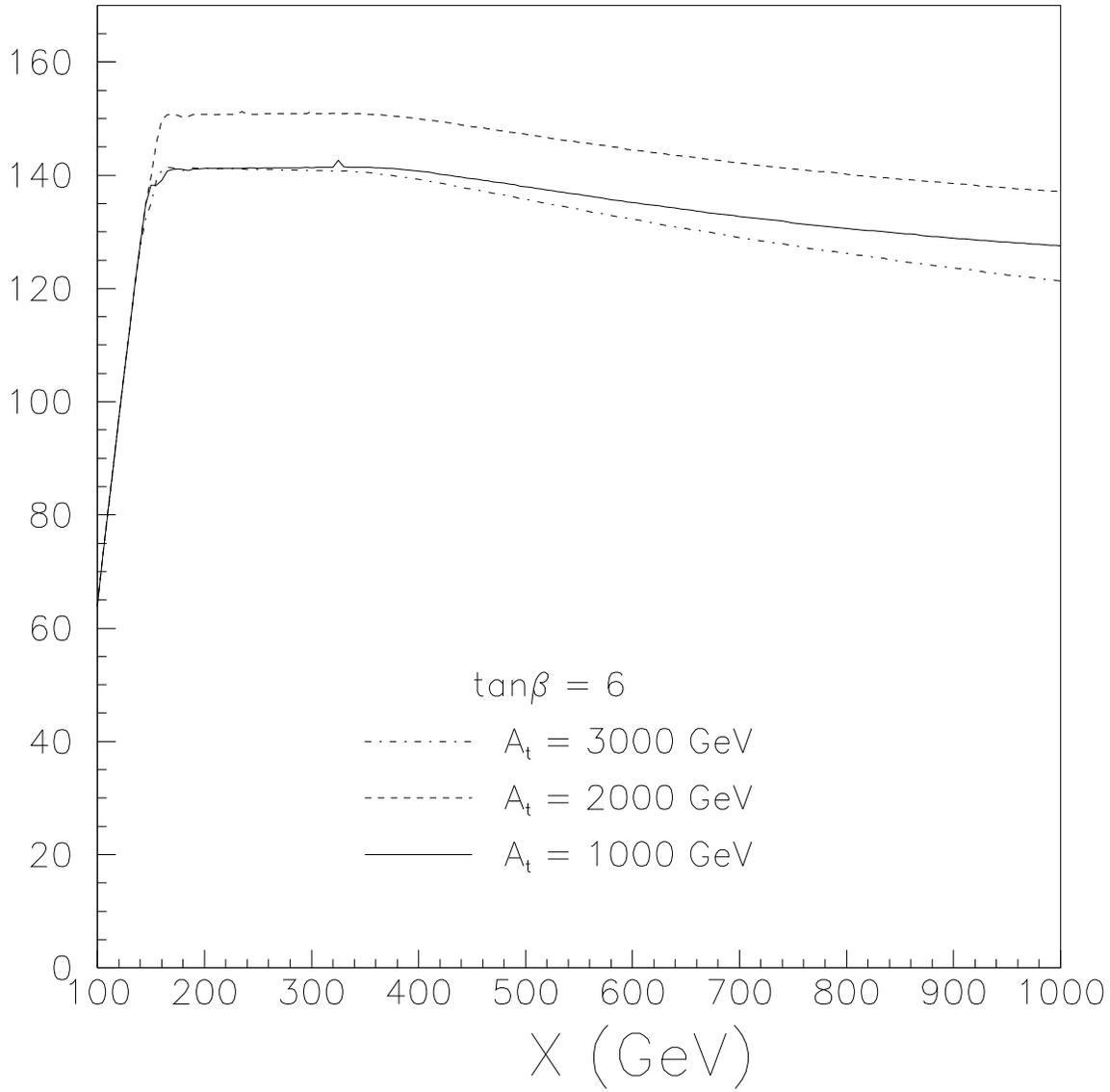
\end{document}